\documentclass[11pt]{article}
\usepackage{amsmath,amsfonts,latexsym,graphicx,amssymb}
\usepackage{amsfonts}
\usepackage{bm}
\usepackage{graphicx,color}

\newcommand{\ra}{{\, \rightarrow\, }}
\newcommand{\ot}{{\,\otimes\,}}
\newcommand{{\Cd}}{{\mathbb{C}^3}}

\def\oper{{\mathchoice{\rm 1\mskip-4mu l}{\rm 1\mskip-4mu l}{\rm 1\mskip-4.5mu l}{\rm 1\mskip-5mu l}}}
\def\<{\langle}
\def\>{\rangle}
\newcommand{\ket}[1]{|#1\rangle}
\newcommand{\bra}[1]{\langle#1|}

\newtheorem{thm}{Theorem}[section]
\newtheorem{proposition}{Proposition}[section]
\newtheorem{definition}{Definition}[section]

\newtheorem{remark}{Remark}[section]
\newtheorem{cor}{Corollary}[section]
\newtheorem{lemma}{Lemma}[section]

\newcommand{\proj}[1]{\ket{#1}\!\bra{#1}}

\begin{document}

\date{}

\title{\textbf{Exposed positive maps: a sufficient condition  }}

\author{Dariusz  Chru\'sci\'nski$^1$ and
Gniewomir Sarbicki$^{1,2}$ \\
$^1$ Institute of Physics, Nicolaus Copernicus University,\\
Grudzi\c{a}dzka 5/7, 87--100 Toru\'n, Poland\\
$^2$ Stockholms Universitet, Fysikum, S-10691 Stockholm, Sweden}

\maketitle

\begin{abstract}
Exposed positive maps in matrix algebras define a dense subset of extremal maps.
We provide a sufficient condition for a positive map to be exposed. This is an analog of a spanning property which guaranties that a positive map is optimal. We analyze a class of decomposable maps for which this condition is also necessary.
\end{abstract}

\section{Introduction}

Positive maps in $\mathbb{C}^*$-algebras play an important role both in mathematics, in connection with the operator theory \cite{Paulsen},
and in modern quantum physics. Normalized positive maps provide an affine mapping between sets of
states of $\mathbb{C}^*$-algebras. In recent years positive maps found important application in entanglement theory defining basic tool for detecting quantum entangled states (see e.g. \cite{HHHH} for the recent review).

Let $\mathfrak{U}$ be a unital $\mathbb{C}^*$-algebra. A linear map $\Phi : \mathfrak{U} \to \mathcal{B}(\mathcal{H})$ is positive if
$ \Phi(\mathfrak{U}_+) \subset \mathcal{B}_+(\mathcal{H})\,$, where $\mathfrak{U}_+$ denotes positive elements in $\mathfrak{U}$.
Denote by $M_k(\mathfrak{U}) = M_k(\mathbb{C}) \ot \mathfrak{U}$ a space of $k \times k$ matrices with entries from $\mathfrak{U}\,$. One says that $\Phi$ is $k$-positive if a linear map $ \Phi_k := \oper_k \ot \Phi : M_k(\mathfrak{U})  \rightarrow M_k(\mathcal{B}(\mathcal{H}))$ is positive. Finally, $\Phi$ is completely positive if it is $k$-positive for $k=1,2,\ldots$. Due to the Stinespring theorem \cite{Stinespring} the structure of completely positive maps is perfectly known: any completely positive map $\Phi$ may be represented in the following form
\begin{equation}\label{CP}
    \Phi(a) = V^\dagger \pi(a) V\ ,
\end{equation}
where $V : \mathcal{H} \rightarrow \mathcal{K}\,$, and $\pi$ is a representation of $\mathfrak{U}$ in the Hilbert space $\mathcal{K}$.
Unfortunately, in spite of the considerable effort, the structure of
positive maps is rather poorly understood \cite{Stormer}--\cite{E}.

Denote by $\mathcal{P}$ a convex cone of positive maps $\Phi : \mathfrak{U} \to \mathcal{B}(\mathcal{H})$. Note, that a space $\mathcal{L}(\mathfrak{U},\mathcal{B}(\mathcal{H}))$ of linear maps from $\mathfrak{U}$ to $\mathcal{B}(\mathcal{H})$ is isomorphic to $\mathcal{B}(\mathcal{H}) \ot \mathfrak{U}\,$. The natural pairing between these two spaces in defined as follows \cite{Eom}: taking an orthonormal basis in  $\cal H$ ($m = {\rm dim}\, \mathcal{H} < \infty$) one identifies $\mathcal{B}(\mathcal{H})$ with $M_m(\mathbb{C})$ and defines
\begin{equation}\label{}
    \< X \ot a, \Phi\> := {\rm tr} (X^{\rm t} \Phi(a) )\ ,
\end{equation}
where $X \in M_m(\mathbb{C})$, $a \in \mathfrak{U}\,$, and $X^{\rm t}$ denotes transposition of $X$ with respect to a given basis in  ${\cal H}$.
Let $\mathcal{P}^\circ$ denote a dual cone \cite{Eom,convex}
\begin{equation}\label{}
  \mathcal{P}^\circ  =   \{\ A \in M_m(C)\otimes \mathfrak{U} \ :\  \<A,\Phi\>\geq 0 \ , \ \Phi \in \mathcal{P}\ \}\ .
\end{equation}
Note that the definition of $\mathcal{P}^\circ$  may be reformulated as follows
\begin{equation}\label{}
    \mathcal{P}^\circ = {\rm conv} \{\  X \ot a \in M_m(\mathbb{C}) \ot \mathfrak{U} \ :\  \< X \ot a,\Phi\> \geq 0 \ , \ \Phi \in \mathcal{P}\ \}\ .
\end{equation}
One finds $\mathcal{P}^\circ = M_m^+(\mathbb{C}) \ot \mathfrak{U}_+\,$,
where $M_m^+(\mathbb{C})$ denotes positive matrices from $M_m(\mathbb{C})$. It shows that $\mathcal{P}^\circ$ defines a convex cone of separable elements in $M_m(\mathbb{C}) \ot \mathfrak{U}\,$.

Recall that a face of $\mathcal{P}$  is a convex subset $F \subset \mathcal{P}$ such that if the convex combination $\Phi = \lambda \Phi_1  + (1-\lambda)\Phi_2$ of $\Phi_1,\Phi_2 \in \mathcal{P}$ belongs to $F$, then both $\Phi_1,\Phi_2 \in F$.
If a ray $\{ \lambda \Phi\, :\, \lambda > 0\}$  is a face of $\cal P$ then it is called an extreme ray, and we say that $\Phi$ generates an extreme ray. For simplicity we call such $\Phi$ an extremal positive map. A face $F$ is exposed if there exists a supporting hyperplane
$H$ for a convex cone $\mathcal{P}$ such that $F=H \cap \mathcal{P}\,$. The property of `being an exposed face' may be reformulated as follows:
A face $F$ of $\mathcal{P}$ is exposed iff  there exists $a\in \mathfrak{U}_+$ and $|h\> \in \mathcal{H}$ such that
$$    F = \{ \ \Phi \in \mathcal{P}\ ; \ \Phi(a)|h\>=0\ \}\ . $$
A positive map $\Phi \in \mathcal{P}$ is exposed if it generates 1-dimensional exposed face.  Let us denote
by ${\rm Ext}(\mathcal{P})$ the set of extremal points and ${\rm Exp}(\mathcal{P})$ the set of exposed points of $\mathcal{P}$. Due to Straszewicz theorem \cite{convex} ${\rm Exp}(\mathcal{P})$  is a dense subset
of ${\rm Ext}(\mathcal{P})$. Thus every extreme map is the limit of some sequence of exposed
maps meaning  that each entangled state may be detected by some exposed positive map.  Hence, a knowledge of exposed maps is crucial for the full characterization of separable/entangled states of bi-partite quantum systems. For recent papers on exposed maps see e.g. \cite{Eom,Marciniak,Majewski,E}.

Now, if $F$ is a face of $\mathcal{P}$ then
\begin{equation}\label{dual_face}
    F' = {\rm conv} \{\ a \ot |h\>\<h| \in \mathcal{P}^\circ\ :\ \Phi(a)|h\> = 0 \ , \ \Phi \in F\ \}\ .
\end{equation}
defines a face of $\mathcal{P}^\circ$ (one calls $F'$ a dual face of $F$). Actually, $F'$ is an exposed face. One proves \cite{Eom}
that a face $F$ is exposed iff $F''=F$.

In this paper we analyze linear positive maps $\Phi : \mathcal{B}(\mathcal{K}) \to \mathcal{B}(\mathcal{H})\,$, where both $\cal K$ and $\cal H$ are finite dimensional Hilbert spaces. We provide a sufficient condition for the map to be exposed. We call it {\em strong spanning property} in analogy to well known spanning property which is sufficient for the map to be optimal \cite{Lew}. Interestingly, this condition is also necessary if $\Phi$ is decomposable and ${\rm dim}\, \mathcal{K} = 2$.  Finally, we characterize the property of exposedness in terms of entanglement witnesses.

\section{Preliminaries}

Consider a positive map $\Phi : \mathfrak{U} \to \mathcal{B}(\mathcal{H})$, where $\mathfrak{U}$ is a unital $\mathbb{C}^*$-algebra and $\mathcal{B}(\mathcal{H})$ denotes a set of bounded operators on the finite dimensional Hilbert space $\mathcal{H}$.

\begin{proposition}
If $a \in \mathfrak{U}$ is strictly positive, i.e. $a \in {\rm int}\, \mathfrak{U}_+$, then ${\rm Im}\,\Phi(b) \subset {\rm Im}\,\Phi(a)$ for all $b  \in \mathfrak{U}_+$.
\end{proposition}

\noindent Proof:  Let us observe that
\begin{equation}\label{}
    {\rm Ker}\, \Phi(a) \subset {\rm Ker}\, \Phi(b)\ .
\end{equation}
Indeed, suppose that there exists $x \in \mathcal{H}$ such that  $x \in {\rm Ker}\, \Phi(a)$  and $ x \not \in {\rm Ker}\, \Phi(b)$. One has $\<x|\Phi(b)|x\> >0$ and $\<x|\Phi(a)|x\>=0$. Now, since $a \in {\rm int}\, \mathfrak{U}_+$ there exists $\epsilon >0$ such that an open ball $B(a,\epsilon) \subset \mathfrak{U}_+$. It is therefore clear that
 \begin{displaymath}
  a' = a - \frac{\epsilon}2 \frac{u-a}{||u-a||}
 \end{displaymath}
belongs to $\mathfrak{U}_+\,$.  One has
\begin{equation}\label{}
    \<x|\Phi(a')|x\> = - \frac{\epsilon}{2||u-a||}\, \<x|\Phi(u)|x\> < 0 \ ,
\end{equation}
which contradicts that $\Phi$ is a positive map. Hence, if $a \in {\rm int}\, \mathfrak{U}_+$, then ${\rm Ker}\,\Phi(a) \subset {\rm Ker} \, \Phi(b)$ for any $b \in \mathfrak{U}_+$ which implies ${\rm Im}\,\Phi(b) \subset {\rm Im}\,\Phi(a)$.
\hfill $\Box$.

\begin{cor}
If $a,b \in {\rm int}\, \mathfrak{U}_+\,$, then ${\rm Im}\,\Phi(a) = {\rm Im}\,\Phi(b)$.
\end{cor}

\begin{cor}
In particular for $a \in  \mathfrak{U}_+\,$ $(a \in {\rm int}\,\mathfrak{U}_+)$, one has ${\rm Im}\,\Phi(a) \subset {\rm Im}\,\Phi(\oper)$ $({\rm Im}\,\Phi(a) = {\rm Im}\,\Phi(\oper))$.
\end{cor}

Let $A:= \Phi(\oper)$. If $A > 0$, that is, $A$ is of full rank, then one has
\begin{equation}\label{Phi-Phi}
    \Phi(a) = A^{1/2} \widetilde{\Phi}(a) A^{1/2}\ ,
\end{equation}
where $\widetilde{\Phi}(a) = A^{-1/2}\Phi(a)A^{-1/2}$ is a unital positive map from $\mathfrak{U}$ into $\mathcal{B}(\mathcal{H})$.  If $A$ is not strictly positive, that is, $A \in \partial \mathcal{B}_+(\mathcal{H})$, then denote by $\mathcal{H}_\Phi$ the range of $A$.
$A$ is invertible on its image and denote by $\widetilde{A}^{-1}$ the generalized inverse of $A$. Now, one has
\begin{equation}\label{Phi-Phi}
    \Phi(a) = A^{1/2}  \widetilde{A}^{-1/2} {\Phi}(a)  \widetilde{A}^{-1/2} A^{1/2}\ .
\end{equation}
Note, that ${\rm Im}\, \Phi(a) \subset \mathcal{H}_\Phi$. Following \cite{Wor2} let us introduce the following

\begin{definition} \label{DEF-EX}
Consider a positive map $\phi : \mathfrak{U} \rightarrow \mathcal{B}(\mathcal{H})$. A map $\phi' : \mathfrak{U} \rightarrow \mathcal{B}(\mathcal{H}')$ is called an {\em extension} of $\phi$ iff $\,\mathcal{H}' \supset \mathcal{H}$ and for any $a \in \mathfrak{U}$
\begin{equation}\label{}
    \phi(a) = \mathcal{P} \phi'(a) \mathcal{P}\ ,
\end{equation}
where $\mathcal{P}$  denotes orthogonal projection $\mathcal{H}' \rightarrow \mathcal{H}$.
\end{definition}
Note that $\mathcal{H}'=\mathcal{H}\oplus\mathcal{H}^\perp$ and hence for any $|h'\> \in \mathcal{H}'$ one has $|h'\> = |h\> \oplus |h^\perp\>$, where $|h\> = \mathcal{P}|h'\>$ which implies
$\phi(a)|h\> = \mathcal{P}\phi'(a)|h\> \,$, and an extension $\Phi'$ is trivial if
\begin{equation}\label{}
    \phi'(a)|h\> = \phi(a)|h\>\ ,
\end{equation}
for all $a \in \mathfrak{U}$ and $|h\> \in \mathcal{H}$.
According to this definition a positive map $\widetilde{\Phi}(a) :=  \widetilde{A}^{-1/2} {\Phi}(a)  \widetilde{A}^{-1/2}$ is a trivial extension of the unital map $\Phi_1 : \mathfrak{U} \to \mathcal{B}(\mathcal{H}_\Phi)$
\begin{equation}\label{}
    {\Phi}_1 = \mathcal{P}_\Phi\, \widetilde{\Phi} \, \mathcal{P}_\Phi\ ,
\end{equation}
where $\mathcal{P}_\Phi$ is a projector $\mathcal{H} \to \mathcal{H}_\Phi$. This way we proved the following

\begin{proposition} Any linear positive map $\Phi : \mathfrak{U} \to \mathcal{B}(\mathcal{H})$ can be written as follows
\begin{equation}\label{}
    \Phi(a) = V^\dagger \Phi_1(a) V\ ,
\end{equation}
where $V : \mathcal{H} \to \mathcal{H}_\Phi$ and $\Phi_1 : \mathfrak{U} \to \mathcal{B}(\mathcal{H}_\Phi)$ is unital.
\end{proposition}
Let us recall

\begin{definition}
A linear map $\Phi$ is irreducible if $[\Phi(a),X] =0$ for all $a \in \mathfrak{U}$ implies that $X = \lambda \mathbb{I}_\mathcal{H}$.
$\Phi$ is irreducible on its image if $[\Phi(a),X] =0$ for all $a \in \mathfrak{U}$ implies that $\mathcal{P}_\Phi X \mathcal{P}_\Phi = \lambda \mathbb{I}_{\mathcal{H}_\Phi}$.
\end{definition}

\begin{remark}
Note, that one may restrict oneself to self-adjoint elements $\mathfrak{U}_{\rm sa}$ only. Indeed, suppose that $\Phi$ is irreducible and $[\Phi(a),X] =0$ for all $a \in \mathfrak{U}_{\rm sa}\,$. Any element $x \in \mathfrak{U}$ may be decomposed as $x = x_1 + i x_2$, with $x_1,x_2 \in \mathfrak{U}_{\rm sa}$. One has
$$ [\Phi(x),X] = [\Phi(x_1),X] + i[\Phi(x_2),X] = 0 \ , $$
and irreducibility of $\Phi$ implies therefore $X = \lambda\mathbb{I}_{\mathcal{H}_\Phi}$.
\end{remark}

\begin{proposition} \label{Xdiag}
 Let a positive map $\Phi$ be irreducible. If $X \Phi(a)=\Phi(a)X^\dagger\,$ for all $a \in \mathfrak{U}\,$,   then $ X=\lambda \mathbb{I}_\mathcal{H}$.
\end{proposition}

\noindent \textbf{Proof:} Irreducibility implies  that $A= \Phi(\oper) > 0$ and hence
\begin{equation}\label{Phi-Phi}
    \Phi(a) = A^{1/2}  {\Phi}_1(a)   A^{1/2}\ ,
\end{equation}
where $\Phi_1$ is unital. One has
$$ X A^{1/2}  {\Phi}_1(a)   A^{1/2}  = A^{1/2}  {\Phi}_1(a)   A^{1/2} X^\dagger\ , $$
and hence
\begin{equation}\label{}
    Y \Phi_1(a) = \Phi_1(a) Y^\dagger\ ,
\end{equation}
with $Y=  A^{-1/2}  X  A^{1/2} $. Using $\Phi_1(\oper) = \mathbb{I}_\mathcal{H}$ one finds $Y^\dagger=Y$. Let us observe that $\Phi_1$ is irreducible as well and hence $Y = \lambda \mathbb{I}_\mathcal{H}$ which implies $X = \lambda \mathbb{I}_\mathcal{H}$. \hfill $\Box$

\section{Exposed maps -- sufficient condition}

In this section we formulate a sufficient condition for a map $\Phi : \mathcal{B}(\mathcal{K}) \longrightarrow \mathcal{B}(\mathcal{H})$ to be exposed. Recall that a linear operator $W \in \mathcal{B}(\mathcal{K}\ot \mathcal{H})$ is block-positive iff $\< x \ot y | W |x \ot y\> \geq 0$ for all product vectors $|x \ot y\> \in \mathcal{K}\ot \mathcal{H}$. Now, due to the Choi-Jamio{\l}kowski isomorphism, $W$ is block-positive iff there exists a positive map $\Phi : \mathcal{B}(\mathcal{K}) \rightarrow \mathcal{B}(\mathcal{H})$ such that
$$ W = (\oper_{\cal K} \ot \Phi) P^+_\mathcal{K}\ , $$
where $\oper_\mathcal{K}$ is an identity map in $\mathcal{B}(\mathcal{K})\,$, and $P^+_\mathcal{K}$ is a maximally entangled state in $\mathcal{K} \ot \mathcal{K}$. Any block-positive but not positive $W$ is called an entanglement witness.
It is therefore clear that any property of a map $\Phi$ may be formulated in terms of $W$ and vice versa.  Now, let us define
\begin{equation}\label{}
    P_W = \{\, x \ot y\ : \ \< x \ot y |W|x \ot y\> =0 \,\}\ .
\end{equation}
Note, that
$$  \< x \ot y |W|x \ot y\> = \<y|\Phi(|\overline{x}\>\<\overline{x}|)|y\> \ , $$
and hence one may equivalently introduce $P_\Phi \equiv P_W = \{\, x \ot y\ : \ \Phi(|\overline{x}\>\<\overline{x}|)|y\> = 0\, \}$. One says that $\Phi$ has {\em spanning property} iff ${\rm span}_\mathbb{C}P_\Phi = \mathcal{K}\ot \mathcal{H}$. Denoting $d_\mathcal{K} = {\rm dim}\, \mathcal{K}$ and $d_\mathcal{H}={\rm dim}\, \mathcal{H}$, one proves \cite{Lew}

\begin{thm}
If a positive map $\Phi$ satisfies spanning property, then it is optimal.
\end{thm}
In analogy we  have the following
\begin{thm}  \label{TH-S}
Let $\Phi :  \mathcal{B}(\mathcal{K}) \longrightarrow \mathcal{B}(\mathcal{H})$ be a positive  map irreducible on its image and
\begin{equation}\label{}
    N_\Phi = {\rm span}_\mathbb{C}\{ \, a \ot |h\> \in \mathcal{B}_+(\mathcal{K}) \ot \mathcal{H} \ : \ \Phi(a)|h\> = 0\, \}\ .
\end{equation}
If the subspace $N_\Phi \subset \mathcal{B}(\mathcal{K}) \ot \mathcal{H}$ satisfies
\begin{equation}\label{STRONG}
{\rm dim}\, N_\Phi = d_\mathcal{K}^2 d_\mathcal{H}-{\rm rank}\,\Phi(\mathbb{I}_\mathcal{K})\ ,
\end{equation}
then $\Phi$ is exposed.
\end{thm}

\noindent \textbf{Proof:} The idea of the proof comes from \cite{Wor2} (see Theorem 3.3). Consider  a map \cite{private}
$$  \tilde{\Phi} \  : \ \mathcal{B}(\mathcal{K}) \ot \mathcal{H} \ \longrightarrow \ \mathcal{H} \  $$
defined by
\begin{equation}
\tilde{\Phi}(a \otimes |h\>) :=  \Phi(a)|h\>\ .
\label{tildePhi}
\end{equation}
Note, that ${\rm dim}\, ({\rm Im}\, \tilde{\Phi}) = {\rm rank}\, \Phi(\mathbb{I}_\mathcal{K})$ and hence $N_\Phi$ defines the kernel of $\tilde{\Phi}$.  To show that $\Phi$ is exposed let us introduce a linear functional $f$ on the space of positive maps $\mathcal{B}(\mathcal{K}) \rightarrow \mathcal{B}(\mathcal{H})$ defined as follows
\begin{equation}\label{f}
 f(\Psi) = \sum_{i=1}^{d_N} \bra{h_i} \Psi(a_i) \ket{h_i}\ ,
\end{equation}
where $d_N$ vectors $a_i \ot |h_i\>$ span $N_\Phi$. Note that $f(\Psi) \geq 0$ for all positive maps $\Psi$ and $f(\Phi)=0$.
As a result $f$ defines a supporting hyperplane to the cone of positive maps $\mathcal{B}(\mathcal{K}) \rightarrow \mathcal{B}(\mathcal{H})$ passing through a map $\Phi$. Note that $\Phi$ is exposed iff $f(\Psi)=0$ implies $\Psi = \lambda \Phi$, with $\lambda$ being a positive number.
Let us observe that  $f(\Psi)=0$  if and only if  $\tilde{\Psi}(a_i \otimes |h_i\>)=\Psi(a_i)|h_i\> = 0\,$, for all $i=1,\ldots,d_N\,$,  and hence the kernel of $\tilde{\Psi}$ contains $N_\Phi$.
To complete the proof we use the following

\begin{lemma} Consider two linear operators $A,B : V \ra W$, where $V$ and $W$ are finite dimensional vector spaces over $\mathbb{C}$.
If $\,{\rm ker} A \supset {\rm ker} B\, $, then there exists $ X: W \ra W$ such that $A=XB\,$ and ${\rm rank}\, X = {\rm rank}\, A$.
\end{lemma}
\noindent \textbf{Proof:} let
$$   A = U_A \Sigma_A V_A^\dagger\ , \ \ \  B = U_B \Sigma_B V_B^\dagger\ , $$
denote the corresponding singular value decompositions of $A$ and $B$. Let $\{ v_\alpha(A)\}$, $\{w_\alpha(A)\}$, $\{ v_\alpha(B)\}$ and $\{w_\alpha(B)\}$ denote the orthonormal basis made from columns of $V_A$, $U_A$, $V_B$, $U_B$ respectively. One has
\begin{equation}\label{}
    \Sigma_A = \sum_{\alpha=1}^{r_A} \sigma_\alpha(A) |w_\alpha(A)\>\<v_\alpha(A)| \ , \ \ \ \
     \Sigma_B = \sum_{\alpha=1}^{r_B} \sigma_\alpha(B) |w_\alpha(B)\>\<v_\alpha(B)| \ ,
\end{equation}
where $ \sigma_\alpha(A)$ and $ \sigma_\alpha(B)$ are strictly positive singular values of $A$ and $B$, respectively. Note, that  condition $\,{\rm ker} A \supset {\rm ker} B, $ is equivalent to $r_B \geq r_A$. One finds $\,A = XB\,$,
where
\begin{equation}\label{}
    X = A V_B^\dagger \widetilde{\Sigma}_B U_B^\dagger\ ,
\end{equation}
with
\begin{equation}\label{}
    \widetilde{\Sigma}_B = \sum_{\alpha=1}^{r_B} \sigma_\alpha(B)^{-1} |w_\alpha\>\<v_\alpha| \ .
\end{equation}
Indeed, one has
$$XB = (A V_B \widetilde{\Sigma}_B U_B^\dagger) ( U_B \Sigma_B V_B^\dagger) = A V_B \widetilde{\Sigma}_B \Sigma_B V_B^\dagger = A \, \sum_{\alpha=1}^{r_B} \proj{v_\alpha(B)}\ . $$ 
Now, since  $\,{\rm ker} A \supset {\rm ker} B\, $,  one has
$$ A \,\sum_{\alpha=1}^{r_B} \proj{v_\alpha(B)} = A\ , $$
which ends the proof.
 \hfill $\Box$

\vspace{.3cm}

One has, therefore,  $\tilde{\Psi}=X \tilde{\Phi}\,$, for some operator $X$ acting on the image of $\tilde{\Phi}$, meaning that
\begin{displaymath}
 \Psi(a) |h\> = X \Phi(a) |h\> \ ,
\end{displaymath}
for all $a \in \mathcal{B}(\mathcal{K})$ and $|h\>\in \mathcal{H}$. Note that for any $a \in \mathcal{B}_{\rm sa}(\mathcal{K})$ one has $\Psi(a) = \Psi(a)^\dagger$ and hence
\begin{equation}\label{hermicity}
  \ X \Phi(a) = \Phi(a) X^\dagger \ .
\end{equation}
Proposition \ref{Xdiag} implies, therefore, that $X \sim \mathbb{I}$ on the image of $\Phi$. Hence $\Psi = \lambda \Phi$ with $\lambda > 0$ due to the fact that both $\Phi$ and $\Psi$ are positive maps.  \hfill $\Box$

\begin{cor} Let $\Phi :  \mathcal{B}(\mathcal{K}) \longrightarrow \mathcal{B}(\mathcal{H})$ be a positive, unital irreducible map. If
\begin{equation}\label{STRONG-I}
    {\rm dim}\, N_\Phi = (d_\mathcal{K}^2 -1)d_\mathcal{H}\ ,
\end{equation}
then $\Phi$ is exposed.
\end{cor}

We propose to call (\ref{STRONG}) {\em strong spanning property} in analogy to spanning poperty
\begin{equation}\label{WEAK}
    {\rm dim}\, {\rm span}_\mathbb{C}\{ |x\> \ot |h\> \in \mathcal{K}\ot \mathcal{H}\ : \ \Phi(|\overline{x}\>\<\overline{x}|)|h\> = 0\, \} = d_\mathcal{K} d_\mathcal{H}\ ,
\end{equation}
which is sufficient for optimality.

\section{A class of exposed decomposable maps $\mathcal{B}(\mathbb{C}^2) \longrightarrow \mathcal{B}(\mathbb{C}^m)$ }

In this section we provide a class of positive exposed maps for which strong spanning property (\ref{STRONG}) is also necessary.

\begin{thm} \label{MAIN}
 Let $\Phi:\mathcal{B}(\mathbb{C}^2) \to \mathcal{B}(\mathbb{C}^m)$ be a decomposable positive but not completely positive map. Then the following conditions are equivalent:
\begin{enumerate}
 \item $\Phi$ is exposed.
 \item $\Phi(\rho)=V^\dagger\rho^{\rm t} V$, where $V: \mathbb{C}^n \to \mathbb{C}^2$ is a linear map of rank two.
 \item There are $4m-2$ linearly independent vectors in the set $\{\, a \otimes |h\> \in \mathcal{B}_+(\mathbb{C}^2) \otimes \mathbb{C}^m \, : \, \Phi(a)|h\>=0\}$.
\end{enumerate}
\end{thm}

\noindent \textbf{Proof:} (1 $\Rightarrow$ 2) Any exposed map is extremal and hence being a decomposable map  $\Phi$ is given by
$\Phi(a)= V^\dagger a V$ or $\Phi(a) = V^\dagger a^{\rm t} V$. The former is evidently CP and the latter in not CP iff ${\rm rank}(V)=2$.

(2 $\Rightarrow$ 3)  Note, that using linear transformation one can transform $V$ to the following form  $V = \sum_{i=1}^2 |e_i\>\<f_i|\,$,  where $\{e_i\}_{i=1}^2, \{f_j\}_{j=1}^m$ are orthonormal bases in $\mathbb{C}^2$ and $\mathbb{C}^m$, respectively. One finds $4(m-2)$ independent vectors taking $a \in \mathcal{B}_+(\mathbb{C}^2)$ arbitrary and $|h\> = \sum_{j=3}^m h_j f_j$. Now, we look for the remaining vectors $a \ot |h\>$, with $|h\> = h_1f_1 + h_2 f_2$.  It is clear that it is enough to consider $a \in \mathcal{B}_+(\mathbb{C}^n)$ being rank-1 projector, i.e. $a = |x\>\<x|$. One has
\begin{equation}\label{}
    \Phi(|x\>\<x|)|h\> = V^\dagger |\overline{x}\>\<\overline{x}|V|h\> = \left(\sum_{i=1}^2 x_ih_i \right)\sum_{j=1}^2 \overline{x}_j |f_j\> \ .
\end{equation}
Note that $ \Phi(|x\>\<x|)|h\> =0$ for $|x\> \neq 0$ if and only if $\sum_{i=1}^2 x_ih_i =0$, and hence (up to trivial scaling) $x_1=h_2$ and $x_2=-h_1$. The family of vectors $|x\>\<x| \ot |h\> \in \mathcal{B}(\mathbb{C}^2) \ot \mathbb{C}^m$  is linearly independent iff the corresponding vectors $|\overline{x}\> \ot |x\> \ot |h\>$ are linearly independent in $\mathbb{C}^2 \ot \mathbb{C}^2 \ot \mathbb{C}^m$.
Note that coordinates of $\bar{x} \otimes x \otimes h$ are polynomial functions of $h_k$ and $\bar{h}_k$, namely:
$$ h_1 h_2 \overline{h}_2 \, , \ h_1 h_2 \overline{h}_1 \, ,\ h_1^2 \overline{h}_2\, ,\ h_1 h_2 \overline{h}_2\, ,\ h_2^2 \overline{h}_2\, ,\ h_2^2 \overline{h}_1\, ,\ h_2 h_1 \overline{h}_2\, ,\ h_2^2 \overline{h}_2 \ . $$
Note, that 6 of them are (functionally) linearly independent and hence one has 6 additional vectors $a \ot |h\>$. Altogether, there are $4(m-2) + 6=4m-2$ linearly independent vectors.

(3 $\Rightarrow$ 1) Follows from Theorem \ref{TH-S}.   \hfill $\Box$

\vspace{.2cm}

A similar problem was analyzed in \cite{Augusiak} in the context of optimal decomposable maps.
Recall that $\Phi$ is decomposable if $\Phi = \Phi_1 + \Phi_2 \circ {\rm t}$, where $\Phi_1$ and $\Phi_2$ are completely positive. Equivalently, the corresponding entanglement witness $W$ is decomposable if $W = Q_1 + (\oper_\mathcal{H} \ot {\rm t})Q_2\,$, where $Q_1,Q_2 \in \mathcal{B}_+(\mathcal{H}\ot \mathcal{K})$.
Let us recall that  $S \subset \mathcal{H} \ot \mathcal{K}$ is a {\em completely entangled subspace} (CES) iff there is no nonzero product vectors in $S$.  The authors of \cite{Augusiak} proved the following

\begin{thm} Let $\Phi  :  \mathcal{B}(\mathbb{C}^2) \longrightarrow \mathcal{B}(\mathbb{C}^m)$ be a positive decomposable map. The following conditions are equivalent

\begin{enumerate}

\item $\Phi$ is optimal,

\item $\Phi(a) = {\rm Tr}_{\mathbb{C}^2} (W\, a^{\rm t} \ot \mathbb{I}_m)\,$, where $W = (\oper_2 \ot {\rm t})Q\,$ and $\,Q\geq 0 $ is supported on a CES,

\item $P_\Phi$ spans $\mathbb{C}^2 \ot \mathbb{C}^m$.

\end{enumerate}
\end{thm}

Note, that we replaced optimality by exposedness, an arbitrary CES by a 1-dimensional CES supporting a positive operator
$$ Q = \sum_{i,j=1}^2 |i\>\<j| \ot V^{\rm t} |i\>\<j| \overline{V} \ , $$
with ${\rm rank}(V)=2$ (clearly, if ${\rm rank}(V)=1\,$, then $Q$ is no longer supported on a CES). Finally, we replaced {\em weak} spanning property
$$ {\rm dim}\, {\rm span}_\mathbb{C}\, \{ \, |\overline{x}\> \ot |h\> \ : \ \Phi(|x\>\<x|)|h\>=0 \, \} = 2m\ , $$
by much stronger property ({\em strong spanning})
$$ {\rm rank}\, \Phi(\mathbb{I}_2) +  {\rm dim}\, {\rm span}_\mathbb{C}\, \{ \, |\overline{x}\> \ot |x\> \ot  |h\> \ : \ \Phi(|x\>\<x|)|h\>=0 \, \} = 4m \ . $$

\section{A class of exposed decomposable maps $\mathcal{B}(\mathbb{C}^n) \longrightarrow \mathcal{B}(\mathbb{C}^m)$ }

It was already shown by Marciniak \cite{Marciniak} that all extremal decomposable maps $\mathcal{B}(\mathbb{C}^n) \longrightarrow \mathcal{B}(\mathbb{C}^m)$ are exposed, i.e. maps of the form $\Phi(a) = V^\dagger aV$ and $\Phi(a) = V^\dagger a^{\rm t} V$ are exposed. Now we show that being exposed these maps in general do not satisfy the {\em strong spanning property} (\ref{STRONG}).

\begin{proposition}
Consider a positive  decomposable map $\Phi : \mathcal{B}(\mathbb{C}^n) \to \mathcal{B}(\mathbb{C}^m)$ defined by $\Phi(a) = V^\dagger a^{\rm t} V$. One has
\begin{equation}\label{}
    {\rm dim}\, N_\Phi = \left\{ \begin{array}{cc} m(n^2-1)  & , \ \ {\rm rank}(V) > 1 \\ mn^2 - (2m-1)    & , \ \ {\rm rank}(V) = 1 \end{array}  \right.
\end{equation}
\end{proposition}

\noindent {\bf Proof:} it is clear that it is enough to consider $a \in \mathcal{B}(\mathbb{C}^n)_+$ being rank-1 projector, i.e. $a = |x\>\<x|$.
Note, that using a linear transformation one can transform $V$ to the following form  $V = \sum_{i=1}^r |e_i\>\<f_i|\,$,  where $\{e_i\}_{i=1}^n, \{f_j\}_{j=1}^m$ are orthonormal bases in $\mathbb{C}^n$ and $\mathbb{C}^m$, respectively.

Let $|\tilde{x}\>$ and $|\tilde{h}\>$ be vectors in $\mathbb{C}^r$ built from the first $r$ coordinates of $|x\>$ and $|h\>$, respectively. For a given vector $|x\>$, the orthogonal complement of $|\tilde{x}\>$ is spanned by $r-1$ vectors
$$ v_2 = |-x_2,x_1,0,\dots,0\>\ ,\ \  v_3=|-x_2,0,x_1,0,\dots,0\>\ ,\  \ldots,\  v_r=|-x_r,\dots,x_1\>\ . $$
 The general vector $|h\>$ orthogonal to $|x\>$ is then of the form $\sum_{i=2}^r \alpha_i |v_i\> \oplus |\hat{h}_i\>$ (where $\sum_{i=2}^r \alpha_i |\hat{h}_i\> = |h_{r+1}, \dots, h_m\>$). Observe, that $|h_{r+1}, \dots, h_m\>$ can be arbitrary. Now, a  general vector $|h\>$ which is orthogonal to $|x\>$ is a linear combination of vectors from $r-1$ subspaces:
\begin{align*}
 & H_2(x) = {\rm span}_\mathbb{C}\{\,|-x_2,x_1, 0, \dots, 0 \>\,\} \oplus \mathbb{C}^{m-r}\ , \\
 & H_3(x) = {\rm span}_\mathbb{C}\{\,|-x_3,0,x_1, 0, \dots, 0  \>\,\}\oplus \mathbb{C}^{m-r} \ , \\
 & \vdots  \\
 & H_r(x) = {\rm span}_\mathbb{C}\{\,|-x_r,0,\dots,0, x_1 \>\,\}\oplus \mathbb{C}^{m-r} \ .
\end{align*}
Consider  the subspace $W_2 \subset \mathbb{C}^n \ot \mathbb{C}^n \ot \mathbb{C}^m \approxeq \mathcal{B}(\mathbb{C}^n) \ot \mathbb{C}^m$  spanned by the vectors $|\overline{x}\> \ot |x\> \otimes |h\>$, where $|h\> \in H_2(x)$, that is, $|\overline{x}\> \ot |y\>\,$, where
\begin{equation} \label{product}
 |y\> =  |x_1, \dots, x_r, x_{r+1}, \dots, x_n\> \otimes |-x_2,x_1,0,\dots,0,h_{r+1},\dots,h_m\> = \sum_{i,j}\, y_{ij}\, e_i \ot f_j\ .
\end{equation}
The coordinates of $|y\>$  are monomials of degree $2$ in variables $\{x_1,\dots,x_n,h_{r+1},\dots,h_m\}$. Note that $|y\>$  has in general $(2+m-r)\times n$ non-zero coordinates, which satisfy one linear condition $y_{11}+y_{22}=0$. Hence ${\rm dim}\, W_2 = n(n[m-r+2]-1)$. Using the same argument one shows that ${\rm dim}\, W_2 = \ldots = {\rm dim}\, W_r\,$. It is easy to show that
\begin{equation}\label{}
    W_i \cap W_j = W_2 \cap \dots \cap W_r\ ,
\end{equation}
for each pair $i \neq j$. Moreover, the constructions of $H_i(x)$ imply
\begin{displaymath}
W_2 \cap \dots \cap W_r = \mathbb{C}^n \otimes (\mathrm{span}_\mathbb{C}\{e_2, \dots, e_n\}\otimes f_1 \oplus \mathbb{C}^n \otimes \mathrm{span}_\mathbb{C}\{f_{r+1}, \dots, f_m\} ) ,
\end{displaymath}
and hence its  dimension equals $n(n-1+[m-r]n)$. Let $W = {\rm span}_\mathbb{C} ( W_2 \cup \ldots \cup W_r)$. One finds
\begin{eqnarray*}
 \dim W &=& \sum_{i=2}^r \dim W_i - (r-2) \cdot \dim (W_2 \cap \dots \cap W_r) \\
  &=& (r-1)n((m-r+2)n-1) - (r-2)n(n-1+(m-r)n) = n^2 m - n\ .
\end{eqnarray*}
Note that if $r=1$, one consider  vectors $|\overline{x}\> \ot |x\> \otimes |h\>$ such that $x_1 h_1 = 0$. Vectors with $x_1=0$ form a $(n-1)^2 m$ dimensional subspace. Vectors  with $h_1=0$ form a $n^2(m-1)$ dimensional subspace. The intersection of these subspaces is $(n-1)^2(m-1)$ dimensional. Finally, one gets $(n-1)^2 m + n^2(m-1) - (n-1)^2(m-1) = n^2m-(2n-1)$ linearly independent vectors.
\hfill $\Box$

It is therefore clear that the {\em strong spanning property}
$$ {\rm rank}\, \Phi(\mathbb{I}_n) +  {\rm dim}\, {\rm span}_\mathbb{C}\, \{ \, |\overline{x}\> \ot |x\> \ot  |h\> \ : \ \Phi(|x\>\<x|)|h\>=0 \, \} = mn^2 \ , $$
supplemented by irreducibility provides only a sufficient condition for exposedness in the same way as {\em weak spanning property}
$$   {\rm dim}\, {\rm span}_\mathbb{C}\, \{ \, |\overline{x}\> \ot |h\> \ : \ \Phi(|x\>\<x|)|h\>=0 \, \} = nm\ , $$
provides only a sufficient condition for optimality. Note, that $\Phi(a) = V^\dagger a^{\rm t} V$ has a {\em strong spanning property} iff ${\rm rank}(V)=n$. However, $\Phi$ is exposed for any $V$ \cite{Marciniak}.

\section{Conclusions}

We have provided a sufficient condition for exposedness  -- {\em strong spanning property} (\ref{STRONG}). It was shown that in the class of decomposable maps $\mathcal{B}(\mathbb{C}^n) \longrightarrow \mathcal{B}(\mathbb{C}^m)$ this condition is also necessary if $n=2$. This result provides an analog of the result of \cite{Augusiak} in the context of optimal maps/witnesses. One calls a block-positive  operator $W \in \mathcal{B}(\mathbb{C}^n \ot \mathbb{C}^m)$ irreducible iff $W$ cannot be written as $W_1 \oplus W_2$, where $W_1$ and $W_2$ are block-positive. One has the following

\begin{proposition}  \label{G-II}
Let $W \in \mathcal{B}(\mathbb{C}^n \ot \mathbb{C}^m)$ be a block-positive irreducible operator. If
\begin{equation}\label{}
    \dim ({\rm Im} \mathrm{Tr}_{\mathbb{C}^n} W) + \dim \{ a \otimes h: \mathrm{Tr}_{\mathbb{C}^n}[W\,a^{\rm t}\otimes \mathbb{I}_m]|h\>=0 \} =n^2m\ ,
\end{equation}
 then $W$ is exposed.
\end{proposition}
If $n=2$, then one proves the following

\begin{proposition} Let $W \in \mathcal{B}(\mathbb{C}^2 \ot \mathbb{C}^m)$ be a block-positive but not positive decomposable operator (i.e. decomposable entanglement witness). The following conditions are equivalent

\begin{enumerate}

\item $W$ is exposed,

\item $W = (\oper_2 \ot {\rm t})Q\,$, and $Q$ is Schmidt rank 2 projector,

\item There are $3m$ linearly independent vectors  $\,|\overline{x}\>\<\overline{x}| \otimes |h\> \in \mathcal{B}_+(\mathbb{C}^2 \otimes \mathbb{C}^n )$ such that $$\< {x}\ot {h}|W|{x}\ot {h}\>=0\ .$$

\end{enumerate}
\end{proposition}

In the forthcoming paper we use the {\em strong spanning property} to analyze exposed positive indecomposable maps.

\section*{Acknowledgments}

It's a pleasure to thank Professor Woronowicz for interesting discussions about exposed and nonextendible maps and Professor Kye for inspiring discussion about facial structure of the cone of positive maps. G.S. was partially supported  by research fellowship within project {\em Enhancing Educational Potential of Nicolaus Copernicus University in the Disciplines of Mathematical and Natural Sciences}   (project no. POKL.04.01.01-00-081/10.)

\end{document}